\documentclass[twocolumn,aps,prl,showpacs,floatfix]{revtex4}
\usepackage{amsmath}
\usepackage{amssymb}
\usepackage{amsfonts}
\usepackage{pifont}
\usepackage{graphicx}%
\newcommand{\bfsymbol}[1]{\mbox{\boldmath $ #1\!$}}
\begin{document}

\title{Ab-initio spin dynamics applied to nanoparticles:
canted magnetism of a finite Co chain along a Pt(111) surface step edge}

\author{B. \'Ujfalussy$^{1}$, B. Lazarovits$^{2}$, L.
Szunyogh$^{2,3}$, G. M. Stocks$^{1}$ and P. Weinberger$^2$}

\affiliation{$^1$Metals and Ceramics Division, Oak Ridge National Laboratory,
Oak Ridge, Tennessee 37831, USA \linebreak
$^2$Center for Computational Materials Science,
Technical University Vienna,
A-1060, Gumpendorferstr. 1.a., Vienna, Austria \linebreak
$^3$Department of Theoretical Physics and Center for Applied
Mathematics and Computational Physics,
Budapest University of Technology and Economics,
Budafoki \'ut 8, H-1521, Budapest, Hungary}

\begin{abstract}
In order to search for the magnetic ground state of surface
nanostructures we extended {\em first principles} adiabatic spin
dynamics to the case of fully relativistic electron scattering.
Our method relies on a constrained density functional theory whereby
the evolution of the orientations of the spin--moments results from a
semi--classical Landau--Lifshitz equation. This approach is applied to
a study of the ground state of a finite Co chain placed along a step edge
of a Pt(111) surface. As far as the ground state spin orientation
is concerned we obtain excellent agreement with the experiment.
Furthermore we observe noncollinearity of the
atom--resolved spin and orbital moments.
In terms of magnetic force theorem calculations we also demonstrate how
a reduction of symmetry leads to the existence of canted magnetic
states.
\end{abstract}
\date{Mar 25, 2004}
\maketitle

Stimulated by the need for ever higher density recording media,
atomic scale magnetic devices are presently at the very focus of
experimental and theoretical research (see, e.g., the
''viewpoint'' drawn by K\"ubler \cite{kubler}). Without doubt,
understanding and design of the relevant physical properties --
magnetic moments, magnetic anisotropy energies, thermal stability,
switching -- of atomic scaled magnets demand detailed knowledge
of their electronic and magnetic structure. Several studies based
on the Hubbard model \cite{ojeda-lopez,uzdin} or on density
functional theory (DFT) \cite{oda,ivanov,hobbs,fujima,anton} have
attempted to explore the, mostly, noncollinear spin ground state
of free and supported metallic clusters. While noncollinearity due
to frustrated exchange interactions can be described in terms of a
nonrelativistic theory \cite{oda,ivanov,hobbs,fujima}, 
general spin structures are subject to an interplay between the
exchange and the spin-orbit interaction and, therefore, can only
be studied within the framework of a relativistic electron theory
\cite{anton}.  

An efficient tool to search for the equilibrium spin arrangement
of a spontaneously magnetized system is to trace the time
evolution of the spin moments until a stationary state is
achieved. The foundations of the so--called first principles
adiabatic spin--dynamics (SD) for itinerant electron systems were
laid by Antropov {\em et al.} \cite{AKH+96} in quite a general
context. In short, for systems with well--defined local (atomic)
moments, the evolution of the time dependent orientational
configuration, $\{ {\bf e}_i(t) \}$, is described by a
microscopic, quasi--classical equation of motion,
\begin{equation} \label{eq:EOM}
\frac{d {\bf e}_i}{dt} = \gamma \: {\bf e}_i \times {\bf B}_i^{eff} +
\lambda \left[ {\bf e}_i \times ({\bf e}_i \times {\bf B}_i^{eff}) \right]
\; ,
\end{equation}
where ${\bf B}_i^{eff}$ is an effective magnetic field averaged
over the cell $i$, $\gamma$ is the gyromagnetic ratio and
$\lambda$ is a damping (Gilbert) parameter. In this equation, the
instantaneous orientational state is evaluated in terms of a
self--consistent calculation within DFT. This formalism was
further developed by Stocks {\em et al.} \cite{SUW+98,UWS+99}
employing a constrained density functional theory\cite{dederichs}.
Here a local (transverse) constraining field, ${\bf B}_i^{con}$ -
that can be determined selfconsistently - ensures the stability
within DFT of the nonequilibrium orientational state demanded by
the equation of motion. Clearly, the internal effective field that
rotates the spins in the absence of a constraint and, therefore,
has to be used in Eq.~(\ref{eq:EOM}) is just the opposite of the
constraining field \cite{SUW+98}.
By merging with the locally selfconsistent multiple scattering
(LSMS) method SD has been applied to study the complex magnetic
orderings in bulk metals and alloys \cite{SUW+98,UWS+99,SSS+02}.

In surface nanostructures, spin-orbit coupling -- its importance
magnified by the reduced, surface symmetry -- obviously plays a
key role in determining magnetic orientations. In order to deal
with exchange splitting and relativistic scattering on an equal
theoretical footing, we use the above first principles constrained
DFT--SD scheme in conjunction with the Kohn--Sham--Dirac equation,
\begin{eqnarray} \label{eq:KSD}
\left[ c \, \mbox{\boldmath $\alpha$} \! \cdot \! {\bf p} +
\beta m c^2 + V({\bf r}) \right. \qquad \qquad \qquad \qquad && \\
\left. + \mu_B \beta \, \mbox{\boldmath $\sigma$} \! \cdot \!
\left( {\bf B}^{xc}({\bf r}) +
{\bf B}^{con}({\bf r}) \right) - E \right] \psi({\bf r}) &=& 0
 \; .  \nonumber
\end{eqnarray}
In this equation, $\mbox{\boldmath $\alpha$}$ and $\beta$ are the
usual Dirac matrices, $\mbox{\boldmath $\sigma$}$ are the Pauli
matrices, $V({\bf r})$ includes the Hartree and
exchange--correlation potentials, while, within the local spin
density approximation (LSDA), ${\bf B}^{xc}({\bf r})$ is an
exchange field interacting only with the spin of the electron. It
should be noted that Eqs.~(\ref{eq:EOM}) and (\ref{eq:KSD})
display the very basics of a relativistic {\em spin--only}
dynamics, inasmuch no attempt is made to trace, distinctly, the
time evolution of the orbital moments. Furthermore, although
implied by use of Eq.~(\ref{eq:KSD}) with Eq.~(\ref{eq:EOM}), we
did not include a torque related to the coupling of the spin-- and
orbital degrees of freedom. As pointed out by Antropov {\em et
al.} \cite{AKH+96} this approach is only applicable when the
deviation between the orientations of the spin and orbital moments
is small.

In order to apply SD to nanostructures of finite size we have
merged the above scheme with the multiple scattering theory (MST)
Green's function embedded cluster method developed by Lazarovits
{\em et al.} \cite{LSW02}. In this, a self--consistent calculation
is first carried out for the surface system in terms of the
relativistic Screened Korringa--Kohn--Rostoker method
\cite{SUW95}. The nanostructure is subsequently {\em inserted}
into this host using the conventional MST embedding formula,
\begin{equation} \label{eq:embed}
\bfsymbol{\tau}^{c}(\epsilon) =\bfsymbol{\tau}^{r}(\epsilon)
\left[\mathbf{I}- \left(\mathbf{t}^{r}(\epsilon)^{-1}-
\mathbf{t}^{c}(\epsilon)^{-1}\right) \bfsymbol{\tau}^{r}(\epsilon)
\right]^{-1} \; .
\end{equation}
In this equation, $\bfsymbol{\tau}^r$($\mathbf{t}^r$) and
$\bfsymbol{\tau}^c$($\mathbf{t}^c$) are site--angular momentum
matrices of the scattering path operators (single--site
t--operators) of the host surface system and the cluster,
respectively, and $\epsilon$ is the energy. By solving this
equation together with the corresponding Poisson equation -- with
appropriate boundary conditions -- a selfconsistent calculation
for the selected cluster can be performed that takes full account
of the environment.


Recently, Gambardella {\em et al.} \cite{GDM+02} reported
the results of experiments on well characterized finite linear
chains of Co atoms located at a step edge of a Pt(111) surface
terrace.  At 45 K the formation of ferromagnetic spin--blocks of
about 15 atoms was observed with an easy magnetization axis normal
to the chain and pointing along a direction of $43^{\rm o}$
towards the step edge. Previous DFT based studies related to this
experiment calculated spin and orbital
moments\cite{KED+02,LSW03,SMO03} and magnetic anisotropy
energies of finite\cite{LSW03} and
infinite\cite{SMO03} chains on Pt(111) surfaces using
a magnetic force theorem (MFT) approach\cite{Jan99}. 
In what follows, we report the first attempt to obtain
this canted magnetic state from first principles in a way that,
simultaneously with the atomic potentials and effective fields,
the directions of the magnetic moments are obtained selfconsistently.

\begin{figure}[htbp]
\begin{center}
\includegraphics[width=0.40\textwidth,clip]{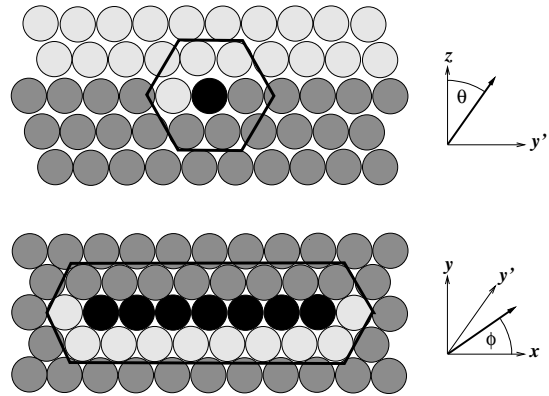} 
\end{center}
\vskip -10pt
\caption{Schematic view of the geometry of a seven--atom Co chain along a Pt(111)
step edge. Full circles: Co atoms,  shaded circles: Pt atoms, open
circles: empty spheres. Top: side view, bottom:
top view of the surface Pt layer with the Co chain.
The embedded cluster is indicated by solid lines.
The coordinate systems
give reference to the azimuthal and polar angles, $\theta$ and
$\phi$, that characterize the orientation of the magnetization.
(Note that in Ref.~\onlinecite{GDM+02} a different coordinate
system and the opposite notation for the angles is used.)}
\label{fig:geometry}
\end{figure}

We first performed a calculation for a Pt(111) surface in which 8
layers of Pt together with 4 layers of vacuum were treated
selfconsistently. 
A seven--atom chain of Co together with 10 empty (vacuum)
spheres were embedded into the topmost Pt layer as schematically
indicated in Fig. \ref{fig:geometry} in order to create a nascent
step edge and nested Co--chain. Simultaneously, all the nearest
neighbors of the Co atoms were re--embedded into the respective Pt
or vacuum layers to allow for relaxation of potentials around the
Co chain. Thus, an effective embedded cluster of a total of 55
atoms was treated selfconsistently. Although the number of Co
atoms we used in our model chain is substantially less than in the
experiment, our previous experience in calculating magnetic
properties of finite chains suggests that the local moments and
the magnetic anisotropy energy contributions of atoms in the
interior of the chain quickly approach the corresponding values of
an infinite chain \cite{LSW03}.

\begin{figure}[htbp]
\begin{center}
\includegraphics[width=0.45\textwidth]{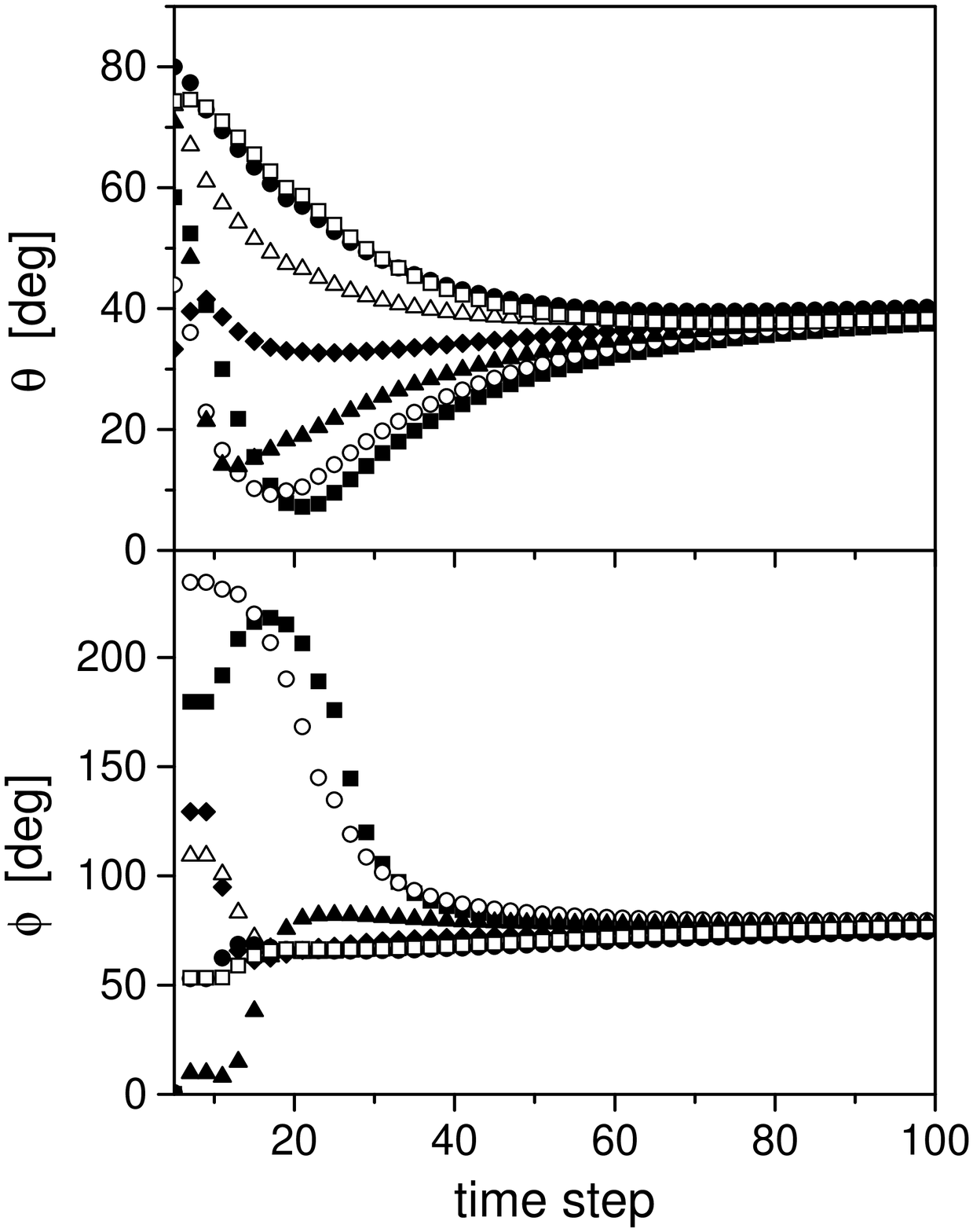} 
\end{center}
\vskip -10pt
\caption{Artificial time evolution of the angles $\theta$ (top)
and the $\phi$ (bottom) angles defining the orientation of the
spin moments for the seven Co atoms in the finite chain depicted
in Fig.~\ref{fig:geometry}. The symbols refer to the following Co
atoms numbered from the left to the right in the bottom part of
Fig.~\ref{fig:geometry}: \ding{110}  \ 1, \ding{109} \ 2,
\ding{115} \ 3, \ding{117} \ 4, $\triangle$ \ 5, \ding{108} \ 6,
$\Box$  \ 7. Shown are the results for only the first 100 time
steps. During the next 900 time steps the angles converge very
smoothly} \label{fig:thph}
\end{figure}

As mentioned before the present implementation of our SD scheme
serves for searching the magnetic ground state of the system. From
this point of view it is sufficient to consider only the second
(damping) term on the right hand side of 
Eq.~(\ref{eq:EOM}). The evolution of the spin orientation is then
measured on a time scale with a unit (time step) of $1/\lambda$. A
stable ground state is signaled by convergence of the $\theta$ and
$\phi$ angles to a constant and concomitant convergence of the
constraining fields to zero. In Fig. \ref{fig:thph} the evolution
of the $\theta$ and $\phi$ angles is plotted for the first 100
steps in this artificial time scale for each Co atom in the chain.
Initially the directions of the atomic magnetic moments were set
by a random number generator. It can be seen that after some
oscillations both the $\theta$ and $\phi$ angles approach a very
similar value for all the Co atoms. This means that the system
rapidly tends to a nearly ferromagnetic order due to the strong
ferromagnetic exchange coupling between the Co atoms. The initial
rapid oscillations seen in Fig.~\ref{fig:thph} are the consequence
of the relatively large constraining fields caused by the large
exchange energies that result whenever the individual site moments
point in very different directions. Once a nearly ferromagnetic
configuration is achieved, the constraining fields drastically
decrease, and a rather slow convergence of the orientations
results. This slow convergence is due to the (much smaller)
spin--orbit interaction energies and it takes about 1000 time
steps for the $\phi$ and $\theta$ angles to converge.
At this point the constraining fields are essentially zero, as is
required for a ground state. The final converged state is
characterized by a $\phi$ angle of $90^\circ$, i.e., normal to the
chain, with a spread of less then $1^\circ$, and $\theta$ angles
of $~42^\circ$. These results are in remarkable agreement with
experiment~\cite{GDM+02}, $\phi=90^\circ,
\theta=43^{\circ}$. It is important to mention that the obtained
ground state is apparently not induced by any symmetry of the
system.

Once such a stable state has been established we can further
analyze the resulting magnetic state in terms of the spin and
orbital moments. For this purpose their size and the corresponding
azimuthal angle $\theta$ for each Co atom are shown in Table
\ref{thetable} for each of the Co atoms in the chain. The first
thing to notice is that the calculated spin moments for the inner
Co atoms (2.18 $\mu_B$) are in good agreement with the value
deduced from experiment (2.12 $\mu_B$)~\cite{GDM+02} and also with
other theoretical studies on infinite wires \cite{KED+02,SMO03}.
Atoms at both ends of the wire have larger spin (and orbital)
moments than those within the wire similarly to our previous
findings\cite{LSW03}. Although our calculated orbital moments for
the inner atoms (0.19--0.20 $\mu_B$) are somewhat larger than the
corresponding values from other LSDA calculations (0.16 $\mu_B$
\cite{KED+02} and 0.15 $\mu_B$ \cite{SMO03}), they are still much
too small when compared to the experimental value (0.68 $\mu_B$)
\cite{GDM+02}. This aspect is a well--known
deficiency of the LSDA and  often  patched up using the so--called
orbital polarization method or the LDA+U method.

\begin{table}[ht]
\begin{tabular}{|c|cc|cc|}
\hline
atom & \multicolumn{2}{c|}{Spin moment} & \multicolumn{2}{c|}{Orbital moment}
\\ \cline{2-5}
  & moment($\mu_B$) & $\Theta$($\deg$) & moment($\mu_B$) & $\Theta$($\deg$) \\\hline
1 & 2.23 & 41.1 & 0.25 & 39.1 \\
2 & 2.18 & 42.5 & 0.20 & 41.5 \\
3 & 2.18 & 42.3 & 0.19 & 40.1 \\
4 & 2.18 & 42.4 & 0.20 & 41.3 \\
5 & 2.18 & 42.3 & 0.19 & 40.2 \\
6 & 2.18 & 42.5 & 0.20 & 41.5 \\
7 & 2.23 & 41.1 & 0.25 & 39.1 \\ \hline
\end{tabular}
\caption{Calculated magnitudes and orientations of the spin and
orbital moments in a seven--atom Co chain along a Pt(111) step edge.}
\label{thetable}
\end{table}

Another interesting feature of the magnetism of finite
nanostructures is the noncollinearity of the moments. As can be
seen from Table \ref{thetable} the spin moments of the inner atoms
are quite parallel while those at the end of the chain are
misaligned by more than 1$^\circ$. This can be traced to larger
anisotropy energy contributions at chain end sites observed in
finite chains earlier\cite{LSW03}.
%
%
Most interestingly Table \ref{thetable} also reveals differences
of as much as 2$^\circ$ between the orientations of respective
spin and orbital moments. This fact underlines the point made by
Jansen \cite{Jan99}, that within DFT the spin and orbital moments
are aligned only when the ground state refers to a high--symmetry
direction.

\begin{figure}[htbp]
\begin{center}
\includegraphics[width=0.45\textwidth,clip]{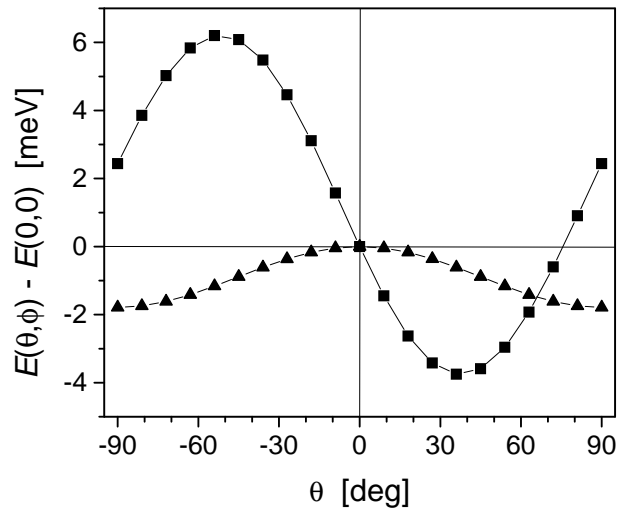}
\end{center}
\vskip -10pt
\caption {Energy curves calculated using the magnetic force
theorem for a ferromagnetic seven--atom Co wire at a Pt(111)
surface step edge as a function of the azimuthal angle, $\theta$.
\ding{110} : $\phi = 90^\circ$ , \ding{115} : $\phi = 0^\circ$.
Solid lines serve as a guide for eyes. } \label{fig:FT}
\end{figure}

While it is necessary to perform fully selfconsistent calculations
to obtain the details of the non-collinear ground state, it is
interesting to study this result in terms of the magnetic force
theorem (MFT) \cite{Jan99}. Assuming ferromagnetic order
great simplification can be achieved by calculating the energy of
the system as a function of the orientation of a by then uniform
magnetization, $E(\theta,\phi)$, such that the effective
potentials and fields are kept fixed. Within this approach only
the single particle (band) energy has to be taken into account.
Fig.~\ref{fig:FT} shows the calculated curves of
$E(\theta,\phi=0^\circ)$ and $E(\theta,\phi=90^\circ)$. In these
calculations we used ground state selfconsistent potentials and
fields obtained from the SD procedure. Clearly, the easy axis
predicted by the MFT calculations, $\theta=38^\circ$ and
$\phi=90^\circ$ is near the one obtained from the SD calculation.
Furthermore, the hard axis is obtained at $\theta=-52^\circ$ and
$\phi=90^\circ$, i.e., by $\Delta \theta = -90^\circ$ away of the
easy axis. This is again in good agreement with
experiment~\cite{GDM+02}.

For surfaces and interfaces with uniaxial symmetry the
easy magnetization axis most often refers to either a normal or a
parallel direction with respect to the planes. 
In case of a $(y,z)$ mirror plane the energy of a ferromagnetic system
can be written up to second order as
\begin{eqnarray}
E(\theta,\phi)&=&E_{0}+K_{2,1}\cos2\theta+ K_{2,2}(1-\cos2\theta)\cos2\phi
 \nonumber \\
&+&K_{2,3}\sin2\theta\sin\phi \; ,
\label{eq:Ethph}
\end{eqnarray}
where $K_{2,i}$ ($i=1,2,3$) are so--called anisotropy constants.
Both calculated trajectories displayed in Fig.~\ref{fig:FT}
coincide almost perfectly with the above function when using the
parameters $K_{2,1}=-0.16$ meV, $K_{2,2}=-1.06$ meV, and
$K_{2,3}=-4.81$ meV. It should be noted that, in terms of these
anisotropy parameters, the easy axis corresponds to a direction,
$\phi=\pi/2$ and
\begin{equation}
\theta = \frac{1}{2} \arctan \left(\frac{K_{2,3}}{K_{2,1}+K_{2,2}} \right) \; .
\end{equation}
The anisotropy energy as defined by the energy difference between
the hard and the easy axes can also be read off Fig.~\ref{fig:FT}:
the corresponding value of 1.42 meV/Co atom compares favorably
with the one deduced from the experiment (2.0 meV) \cite{GDM+02}.
It is worthwhile to mention, that in terms of similar MFT calculations
Shick {\em at al.} obtained an easy axis along $\theta=18^\circ$ and
an anisotropy energy of 4.45 meV \cite{SMO03}. The difference of these
values with respect to the present results can most possibly accounted
for in the film geometry and/or the second--variational treatment of the
spin--orbit coupling used in the FLAPW calculations of Ref.\cite{SMO03}.

In this study we presented the first application of a relativistic
ab inito spin dynamics as based on a constrained density
functional theory to finite magnetic nanostructures. In excellent
quantitative agreement with the corresponding experiment, we
obtained a canted ground state for a Co wire along a Pt(111)
surface step edge. We also found that this magnetic state is
noncollinear: a feature that is expected to play a key role in
nanostructures having complex geometry. For the present relatively
simple geometry the magnetic force theorem proved to be a useful
tool to interpret the results of spin dynamics calculations and
providing additional information such as anisotropy constants and
the anisotropy energy.

Financial support was provided by the Center for Computational
Materials Science (Contract No. GZ 45.451), the Austrian Science
Foundation (Contract No. W004), and the Hungarian National
Scientific Research Foundation (OTKA T038162 and OTKA T037856).
The work of BU and GMS was supported by DOE-OS, BES-DMSE under
contract number DE-AC05-00OR22725 with UT-Battelle LLC.
Calculations were performed at ORNL-CCS (supported by OASCR-MICS)
and NERSC (supported by BES-DMSE).


\begin{thebibliography}{30}

\bibitem{kubler}
J. K\"ubler, J. Phys.: Condens. Matter {\bf 15}, V21 (2003)

\bibitem{ojeda-lopez}
M.~A. Ojeda-L\'opez, J. Dorantes-D\'avila, and G.~M. Pastor,
J. Appl. Phys. {\bf 81}, 4170 (1997).

\bibitem{uzdin}
S. Uzdin, V. Uzdin, and C. Demangeat,
Comp. Mat. Sci. {\bf 17}, 441 (2000); Surf. Sci. {\bf 482-485}, 965
(2001).

\bibitem{oda}
T. Oda, A. Pasquarello, and R. Car,
Phys. Rev. Lett. {\bf 80}, 3622 (1998)

\bibitem{ivanov}
O. Ivanov and V.P. Antropov, J. Appl. Phys. {\bf 85}, 4821 (1999).

\bibitem{hobbs}
D. Hobbs, G. Kresse, and J. Hafner,
Phys. Rev. B {\bf 62}, 11556 (2000).

\bibitem{fujima}
N. Fujima, Eur. Phys. J. D {\bf 16}, 185 (2001).

\bibitem{anton}
J. Anton, B. Fricke, and E. Engel,
Phys. Rev. A {\bf 69}, 012505 (2004).




\bibitem{AKH+96}
V.~P. Antropov, M.~I. Katsnelson, B.~N. Harmon, M.~van Schilfgaarde, and D. Kusnezov,
Phys. Rev. B {\bf 54}, 1019 (1996).

\bibitem{SUW+98}
G.~M. Stocks, B.~\'Ujfalussy, Xindong Wang, D.~M.~C. Nicholson, W.~A.  Shelton, Yang Wang, A. Canning and B.~L. Gy\"orffy,
Philos. Mag. B {\bf 78}, 665 (1998).

\bibitem{UWS+99}
B. \'Ujfalussy, Xindong Wang, D.~M.~C. Nicholson, W.~A.  Shelton, G.~M. Stocks,  Yang Wang, and  B.~L. Gy\"orffy,
J. Appl. Phys.  {\bf 85}, 4824 (1999).

\bibitem{dederichs}
P.~H. Dederichs, S. Bl\"{u}gel, R. Zeller and H. Akai,
Phys. Rev. Lett. {\bf 53}, 2512, (1984).

\bibitem{SSS+02}
G.~M. Stocks, W.~A. Shelton, T.~C. Schulthess, B. \'Ujfalussy, W.~H. Butler, and A. Canning,
J. Appl. Phys. {\bf 91}, 7355 (2002).

\bibitem{LSW02}
B. Lazarovits, L. Szunyogh, and P. Weinberger,
Phys. Rev. B {\bf 65}, 104441 (2002).

\bibitem{SUW95}
L. Szunyogh, B. \'Ujfalussy, and P. Weinberger,
Phys. Rev. B {\bf 51}, 9552 (1995).

\bibitem{GDM+02}
P. Gambardella, A. Dallmeyer, K. Maiti, M.~C. Malagoli, W. Eberhardt, K. Kern, and C. Carbone,
Nature {\bf 416},  301  (2002);
P. Gambardella, J. Phys.: Condens. Matter {\bf 15}, S2533 (2003).

\bibitem{KED+02}
M. Komelj, C. Ederer, J.~W. Davenport, and M. F\"ahnle,
Phys. Rev. B {\bf 66}, 140407 (2002).

\bibitem{LSW03}
B. Lazarovits, L. Szunyogh, P. Weinberger,
Phys. Rev. B {\bf 67}, 024415 (2003).

\bibitem{SMO03}
A.~B. Shick, F.~M\'aca, and P.~M. Oppeneer,
{\tt arXiv:cond-mat/0312467}

\bibitem{Jan99}
H.~J. Jansen, Phys. Rev. B {\bf 59},  4699  (1999).

\end{thebibliography}
\end{document}